\documentclass{article}
\usepackage[T1]{fontenc}
\usepackage[utf8]{inputenc}
\usepackage[]{ismir} 
\usepackage{amsmath, cite, url}
\usepackage{graphicx}
\usepackage{color}

\usepackage{url}
\usepackage{graphicx}

\usepackage{inconsolata}
\usepackage{array}
\usepackage{pifont}
\usepackage{tabularx}
\usepackage{adjustbox}
\usepackage{multirow}
\usepackage{enumitem}
\usepackage{xspace}
\usepackage{tcolorbox}
\usepackage{booktabs,subcaption,amsfonts,dcolumn}
\usepackage{url}
\usepackage{amsmath,amsthm,amsfonts,amssymb,bm,stmaryrd, bbm}
\usepackage{color,xcolor,colortbl}
\usepackage{CJKutf8}
\usepackage{makecell}
\usepackage{booktabs}
\usepackage{graphicx}
\usepackage{listings}
\usepackage{float}
\usepackage{wrapfig}
\usepackage{rotating}
\usepackage{amssymb}
\usepackage{wasysym}

\definecolor{carminepink}{rgb}{0.92, 0.3, 0.26}
\definecolor{jade}{rgb}{0.0, 0.66, 0.42}

\usepackage{pifont}

\definecolor{warning}{HTML}{C80000}
\definecolor{mygreen}{rgb}{0.64, 0.76, 0.68}
\definecolor{myyellow}{rgb}{0.98, 0.94, 0.75}
\definecolor{mygreen}{rgb}{0.68, 0.85, 0.9}
\definecolor{myblue}{rgb}{0.82, 0.94, 0.75}
\definecolor{mypurple}{RGB}{224, 65, 245}
\definecolor{myorange}{RGB}{209, 136, 17}
\definecolor{Mycolor1}{HTML}{BAD8F2}
\definecolor{Mycolor2}{HTML}{DDEEFA}
\definecolor{carminepink}{rgb}{0.92, 0.3, 0.26}
\definecolor{jade}{rgb}{0.0, 0.66, 0.42}

\definecolor{Mycolor1}{HTML}{BAD8F2}
\definecolor{Mycolor2}{HTML}{DDEEFA}

\usepackage{array}
\newcolumntype{Y}{>{\centering\arraybackslash}X}

\title{Evaluating Music Context Preservation:\\A Multi-Facet Framework for Music Editing Systems}

\multauthor
  {Yash Vishe \hspace{1cm} Eric Xue \hspace{1cm} Xunyi Jiang}
  {{\bf  Zachary Novack \hspace{1cm} Junda Wu \hspace{1cm} Julian McAuley \hspace{1cm} Xin Xu}\\
  University of California, San Diego\\
  {\tt\small \{yvishe, exue, xuj003, znovack, juw069, jmcauley, xinxucs\}@ucsd.edu}
  }

\def\authorname{Y. Vish, E. Xue, X. Jiang, Z. Novack, J. Wu, J. McAuley, and X. Xu}

\usepackage[bookmarks=false,pdfauthor={\authorname},pdfsubject={\pdfsubject},hidelinks]{hyperref}

\begin{document}

\maketitle

\begin{abstract}
Music editing plays a vital role in modern music production, with applications in film, broadcasting, and game development.
Recent advances in music editing systems have enabled diverse editing tasks such as timbre transfer, instrument substitution, and genre transformation. 
However, many existing works overlook evaluating their ability to preserve musical facets that should remain unchanged during editing, which we define as Music Context Preservation (MuseCP). 
While some studies do consider MuseCP, their evaluation protocols and metrics are not comprehensive.
To address this, we introduce the first MuseCP evaluation framework, \textbf{MuseCPEval}, that covers four categories of music facets with fine-grained and well-tailored metrics to capture nuanced changes in music attributes.
Objective validation and a human study demonstrate the effectiveness of these metrics.
Moreover, the case studies on diverse music editing systems illustrate the practical utility of these metrics as a testbed and diagnostic tool, providing insights into the strengths and limitations of existing systems.
We hope our metrics and findings can offer practical guidance for developing more effective and reliable music editing strategies with strong MuseCP capability\footnote{Code: \url{https://github.com/Yashvishe13/MuseCPEval}, Demo: \url{https://musecpeval-demo.netlify.app/}}.
\end{abstract}

\section{Introduction}\label{sec:introduction}
\begin{figure}
    \centering
    \includegraphics[width=0.9\linewidth]{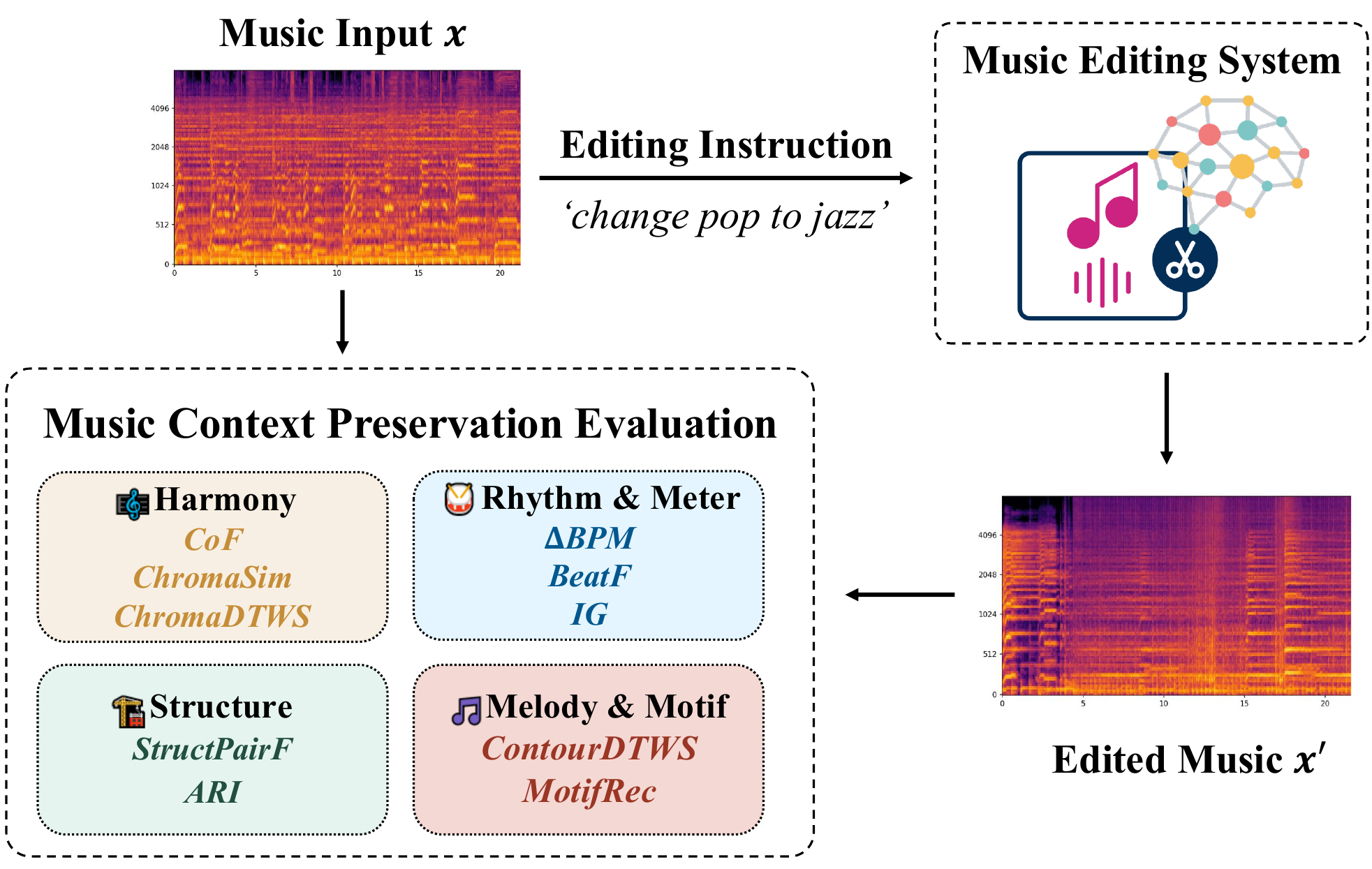}
    \caption{Music context preservation evaluation overview.}
    \label{fig:overview}
\end{figure}
Music editing has emerged as an important direction in recent research and real-world music applications \cite{melodyflow,editingmusicwithmelodyandtext, ramoneda2024musicproofreadingrefinpaintmodify,notthatgroove,mundada-etal-2025-wildscore, wang2025csymr}.
Although recent approaches have achieved promising progress in these areas, their evaluation for \textbf{Music Context Preservation} (MuseCP) capability remains limited in scope or even missing, which we define as the ability of a music editing system to retain the fundamental musical attributes of the source material that are not intended to be edited.
These attributes constitute essential facets of music theory, such as harmony, rhythm, and structure, and their persistence during editing is therefore highly significant and warrants thorough evaluation.
However, as shown in Table \ref{tab:baselines}, the MuseCP evaluation protocols of existing music editing works are not comprehensive.
For instance, ZETA \cite{manor2024zeroshotunsupervisedtextbasedaudio}, Audio Prompt Adapter \cite{audiopromptadapter}, and SteerMusic \cite{steermusic} only evaluated structure, harmony, and melody, respectively.
Even some work didn't consider MuseCP evaluation.
This lack of well-rounded MuseCP evaluation limits our understanding of the strengths and weaknesses of current music editing systems.
Meanwhile, to our best knowledge, there is no framework or tool for systematically measuring the MuseCP capability of music editing systems.

To address this issue, we construct \textbf{MuseCPEval} (Figure \ref{fig:overview}), the first comprehensive MuseCP evaluation framework for music editing systems, which covers four music facets with carefully tailored fine-grained metrics.
Overall, here are the main contributions of our work:

\begin{table*}[htbp]
  \centering
  \scalebox{0.74}{
    \begin{tabular}{lcccc}
    \toprule
    \textbf{Method} & \textbf{Backbone} & \textbf{Task} & \textbf{MuseCP Evaluation?} & \textbf{Evaluated Facet} \\
    \midrule
    \textbf{AUDIT} \cite{wang2023audit} & Diffusion & Add, Drop, Replace, Inpaint, Super-resolution & \color{carminepink}\ding{55}    & ---  \\
    \textbf{InstructME} \cite{instructme} & Diffusion  & Add, Remove, Extract, Replace, Remix & \color{jade}\ding{51}   & Harmony, Rhythm \\
    \textbf{MusicMagus} \cite{zhang2024musicmaguszeroshottexttomusicediting} & Diffusion  & Music Attribute Modification & \color{jade}\ding{51}   & Harmony, Structure \\
    \textbf{ZETA} \cite{manor2024zeroshotunsupervisedtextbasedaudio} & Diffusion  & Music Attribute Modification & \color{jade}\ding{51}   & Structure \\
    \textbf{Audio Prompt Adapter} \cite{audiopromptadapter} & Diffusion  & Timbre and Genre Transfer & \color{jade}\ding{51}   & Harmony \\
    \textbf{Instruct-MusicGen} \cite{instructmusicgen} & Transformer  & Add, Separate, and Remove Stems & \color{carminepink}\ding{55}    & --- \\
    \textbf{Melodia} \cite{melodia} & Diffusion  & Music Attribute Modification & \color{jade}\ding{51}   & Harmony, Melody, Structure \\
    \textbf{SteerMusic} \cite{steermusic} & Diffusion  & Music Attribute Modification & \color{jade}\ding{51}   & Melody \\
    \bottomrule
    \end{tabular}}
    \caption{MuseCP evaluation of recent music editing systems (sorted by publication year, 2023-2026).}
  \label{tab:baselines}%
\end{table*}%

\begin{enumerate} [itemsep=1pt, topsep=2pt, leftmargin=13pt]
    \item \textbf{Introducing MuseCPEval (\S \ref{sec:MuseCP})}: We propose the first MuseCP evaluation framework MuseCPEval. It reorganizes the music facets that are commonly considered to be preserved into four categories: harmony, rhythm \& meter, structure, and melody \& motif.  For each facet, fine-grained metrics are tailored and collected.

    \item \textbf{Validating MuseCP evaluation metrics (\S \ref{sec:validation})}: The metrics are validated through both objective experiments and human listening studies, demonstrating their effectiveness in capturing nuanced music attribute changes and strong alignment between metric values and human perception of musical differences.
    
    \item \textbf{Conducting case studies on diverse music editing systems (\S \ref{sec:setups})}: We apply MuseCPEval to four representative music editing systems spanning diverse methodologies, model architectures, and editing tasks. Through detailed analyses of their strengths and limitations in preserving different music attributes, we show that MuseCPEval serves as a practical testbed and diagnostic tool, offering insights for developing more reliable music editing systems.
    
\end{enumerate}

\section{MuseCPEval}
\label{sec:MuseCP}
\subsection{Music Context Preservation Evaluation}
For most existing music systems, given an original audio music input $x$ and a user's editing instruction, they will produce an edited music $x'$, where certain music attributes are expected to be edited while all other contexts should remain preserved.
In this work, we introduce a framework for evaluating the extent to which non-target music attributes are preserved under editing.
We reorganize widely considered music attributes into four music facets: \textbf{Harmony, Rhythm \& Meter, Structure, and Melody \& Motifs}, covering fundamental dimensions of music \cite{kader2025surveyevaluationmetricsmusic}.
For each facet, we define fine-grained metrics that measure different aspects of similarity between the reference $x$ and the edited output $x'$ within the facet.
The overall evaluation thus provides a multi-facet view of MuseCP.
In this section, we introduce these metrics in detail.

\subsection{Harmony}
Harmony is an important concept in music, representing the basic color of the music. We measure this across high-level keys and scales to low-level chords and chroma using three metrics.

\textbf{Circle-of-fifths (CoF) Distance.}\quad
As for key comparison, we use CoF Distance \cite{DBLP:journals/tmm/ZhuK06} to measure tonal relatedness between $x$ and $x'$.
The tonic $k \in \{0,1,\ldots,11\}$ (12 pitch classes: C, C$^\sharp, \dots$, B) and mode (major or minor) of $x$ and $x'$ ($k'$) are estimated via the Krumhansl–Schmuckler algorithm \cite{10.1093/acprof:oso/9780195148367.001.0001}.
To account for the perceptual equivalence of a minor key and its relative major, each minor tonic is normalized to its relative major via $\tilde{k} = (k + 3) \bmod 12$, while major tonics are left unchanged.
Each $\tilde{k}$ is then mapped to a position $\phi(\tilde{k}) \in \{0,1,\ldots,11\}$ on the circle of fifths, where adjacent positions differ by a perfect fifth.
The distance is then calculated and normalized to $[0,1]$ as:
\begin{equation}
\small
    \text{CoF}(x, x') = \frac{1}{6}\min\left(\left|\phi(\tilde{k}) - \phi(\tilde{k}')\right|,\ 12 - \left|\phi(\tilde{k}) - \phi(\tilde{k}')\right|\right)
\label{eq1}
\end{equation}
A score of 0 means the $x$ and $x'$ share the same key signature; A score of 1 indicates maximally distant keys (e.g., C major vs. F$^\sharp$ major).

\textbf{Chroma Similarity.}\quad
Chroma Similarity \cite{zhang2024musicmaguszeroshottexttomusicediting, audiopromptadapter} is used to capture the chroma distribution similarity between $x$ and $x'$. 
Concretely, $x$ and $x'$ are each converted into a chroma matrix $\mathbf{H} \in \mathbb{R}^{T \times 12}$ via Constant-Q Transform \cite{CQT}, where $T$ is the number of frames and the 12 dimensions correspond to 12 chromas. 
The mean chroma vector $\bar{\mathbf{h}} \in \mathbb{R}^{12}$ is obtained by averaging $\mathbf{H}$ over the time axis, representing the global chroma energy distribution of the music. 
Chroma Similarity is then the cosine similarity between $\bar{\mathbf{h}}$ and $\bar{\mathbf{h}}'$, the mean chroma vectors of $x$ and $x'$ respectively:
\begin{equation}
\small
\text{ChromaSim}(x, x') = \frac{\bar{\mathbf{h}} \cdot \bar{\mathbf{h}}'}{|\bar{\mathbf{h}}| |\bar{\mathbf{h}}'|}
\end{equation}
It ranges in $[0, 1]$, where 1 indicates identical chroma distributions and 0 indicates completely dissimilar.

\textbf{Chroma DTW Similarity.}\quad
To further measure the temporal dynamics of pitch-class content, we additionally compute Chroma Dynamic Time Warping (DTW) Similarity \cite{dtw}.
We compute a pairwise cosine distance matrix $\mathbf{D} \in \mathbb{R}^{T \times T'}$, where $\mathbf{h}_i \in \mathbb{R}^{12}$ and $\mathbf{h}_j' \in \mathbb{R}^{12}$ denote the $i$-th$_{i=1}^T$ and $j$-th$_{j=1}^{T'}$ frame of $\mathbf{H}$ and $\mathbf{H}'$ respectively, and $D_{ij} = 1 - \frac{\mathbf{h}_i \cdot \mathbf{h}_j'}{|\mathbf{h}_i||\mathbf{h}_j'|}$. 
DTW finds the optimal path $P^* = \arg\min\limits_{P} \sum_{(i,j)\in P} D_{ij}$, where $P$ is the set of all valid monotonic and continuous index-pair sequences from $(1,1)$ to $(T, T')$.
The Chroma DTW Similarity is defined as the mean cosine similarity along the optimal path:
\begin{equation}
    \text{ChromaDTWS}(x,x')
= \frac{1}{|P^*|} \sum_{(i,j)\in P^*} (1 - D_{ij})
\end{equation}
It's in $[0,1]$ where 1 means the two chroma sequences are identical and 0 means completely dissimilar.


\subsection{Rhythm \& Meter}
Rhythm and meter describe the temporal organization of music. We evaluate their preservation between $x$ and $x'$ using three metrics at different levels of granularity.

\textbf{Folded Beat Difference.}\quad
This measures the global tempo difference.
The tempos of $x$ and $x'$ are estimated in beats-per-minute (BPM) via \textit{librosa} \cite{librosa} beat tracker, as $\text{BPM}$ and $\text{BPM}'$ respectively.
The absolute BPM difference is then computed, but naively comparing BPM values is susceptible to octave errors, where the beat tracker may estimate the tempo of the same music as $\text{BPM}$, $2\times\text{BPM}$, or $\frac{1}{2}\times\text{BPM}$ depending on the metric level it locks onto. 
To handle this, we fold the estimate by taking the minimum difference folded over the metric octave:
\begin{equation}
\scalebox{0.77}{$
\Delta_{\text{BPM}}(x, x') = \min\left(|\text{BPM} - \text{BPM}'|,\ |\text{BPM} - 2\cdot\text{BPM}'|,\ \left|\text{BPM} - \frac{\text{BPM}'}{2}\right|\right)$}
\label{eq4}
\end{equation}
A value of 0 indicates identical tempo; larger values indicate greater tempo deviation.

\textbf{Beat F-measure.} \quad
It assesses whether beats fall in the right places. Beat positions are first extracted from $x$ and $x'$ as a sequence of timestamps in seconds. 
A beat in $x'$ is counted as a true positive if it falls within a tolerance window ($\pm 70$ms) around a reference beat in $x$. The F-measure is then:
\begin{equation}
    \text{BeatF}(x, x') = \frac{2 \cdot \text{P} \cdot \text{R}}{\text{P} + \text{R}}
\end{equation}
where $\text{P}$ and $\text{R}$ are the precision and recall of matched beats under the \textit{mir\_eval} framework \cite{Raffel2014MIR_EVALAT}. It ranges in $[0, 1]$, where 1 means every beat in $x'$ perfectly aligns with a beat in $x$, and 0 means no correct hits.

\textbf{Information Gain (IG).} \quad
We adopt the beat information gain from \textit{mir\_eval} to measure how consistently the beat phase of $x'$ aligns with that of $x$. For each reference beat, we compute its signed timing error relative to the nearest estimated beat, normalized by the local beat period.
The procedure is repeated in the reverse direction to penalize both missed and spurious beats. 
All normalized errors are binned into a circular histogram of $K=41$ bins, yielding an empirical distribution $p$. The IG is then defined as the KL divergence from $p$ to the uniform distribution:
\begin{equation}
    \text{IG}(x, x') = (\log_2 K - H(p))/\log_2 K
\end{equation}
where $H(p) = -\sum_{k=1}^{K} p_k \log_2 p_k$ is the entropy of $p$. It ranges in $[0,1]$. A high score indicates that timing errors concentrate at a consistent phase offset, reflecting a stable, regular pulse, while a low score indicates that errors are spread uniformly, implying no coherent metrical relationship between $x$ and $x'$.


\subsection{Structure}
Structure describes the high-level organization of music, such as how it is divided into sections and how those sections relate to one another.
We evaluate its preservation by using two complementary metrics. 

\textbf{Structural Pairwise F-measure.}\quad
Music segments are extracted from $x$ and $x'$ via \textit{msaf} \cite{msaf}. Each segment spans a set of time points, where each time point inherits the label of the segment it belongs to. Given the reference labeling from $x$ and the estimated labeling from $x'$, we examine all pairs of time points and ask whether each pair shares a label in both labelings. A true positive occurs when both labelings agree that a pair belongs to the same segment; a false positive when only $x'$ groups them together; a false negative when only $x$ does. 
The precision $P_{\text{pair}}$ and recall $R_{\text{pair}}$ are the true-positive rate with respect to $x'$ and $x$, respectively. 
The Structural Pairwise F-measure is:
\begin{equation}
    \text{StructPairF}(x, x') = \frac{2 \cdot P_{\text{pair}} \cdot R_{\text{pair}}}{P_{\text{pair}} + R_{\text{pair}}} \tag{7}
\end{equation}
It ranges in $[0, 1]$, where 1 means the two labelings induce identical pairwise groupings and 0 means no agreement. 

\textbf{Adjusted Rand Index (ARI).}\quad 
The Rand Index additionally accounts for pairs that both labeling assign to different segments, i.e., agreement on both togetherness and separation \cite{6fd45ff5-8145-3f39-a258-8e3ef378c6a4}. ARI further corrects for chance by subtracting the expected agreement under random labeling:
\begin{equation}
    \text{ARI}(x, x') = \frac{\text{RI} - \mathbb{E}[\text{RI}]}{\max(\text{RI}) - \mathbb{E}[\text{RI}]} \tag{8}
\end{equation}
where RI is the fraction of all time-point pairs on which the two labelings agree. 
A score of 1 indicates perfect agreement, 0 indicates chance-level agreement, and negative values (rare in practice) indicate worse-than-chance clustering. By additionally rewarding agreement on separation and discounting accidental matches, ARI provides a more comprehensive view of structural preservation.

\subsection{Melody \& Motif}
Melody and its recurring short motifs are among the most salient cues by which listeners recognize a piece.
Since this facet focuses solely on melodic content, we first isolate it from key-level variation and then use the following two metrics to measure its preservation.

\textbf{Contour DTW Similarity.}\quad
This metric captures whether the pitch-class content of the piece is preserved \cite{DBLP:conf/ismir/BittnerSEB15}. 
We characterize frame-level pitch content with 12-dimensional CQT chromagrams \cite{Brown1991CalculationOA}, yielding per-frame pitch-class sequences $\mathbf{d} \in \mathbb{R}^{T \times 12}$ and $\mathbf{d}' \in \mathbb{R}^{T' \times 12}$, whose rows $\mathbf{d}_i$ and $\mathbf{d}_j'$ encode the pitch-class distribution of a single frame.
We then compute a pairwise cosine distance matrix $\mathbf{D} \in \mathbb{R}^{T \times T'}$ with $D_{ij} = 1 - \frac{\mathbf{d}_i \cdot \mathbf{d}'_j}{|\mathbf{d}_i||\mathbf{d}'_j|}$.
DTW finds the optimal warping path $P^* = \arg\min\limits_{P} \sum_{(i,j)\in P} D_{ij}$, where $P$ is the set of all valid monotonic and continuous index-pair sequences from $(1, 1)$ to $(T, T')$. The Contour DTW Similarity is defined as the mean cosine similarity along the optimal path:
\begin{equation}
   \text{ContourDTWS}(x, x') = \frac{1}{|P^*|} \sum_{(i,j)\in P^*} (1 - D_{ij})
\end{equation}
It clips to $[0, 1]$, where 1 indicates identical pitch-class trajectories up to tempo-elastic alignment. Lower values indicate greater divergence in pitch-class content.

\begin{table*}[htbp]
  \centering
  \scalebox{0.58}{
    \begin{tabular}{l|cccccc|cccc}
    \toprule
    \textbf{Music Facets} & \multicolumn{6}{c}{\textbf{Harmony}}                   & \multicolumn{4}{c}{\textbf{Structure}} \\
    \midrule
    \textbf{Editing}  & \textbf{CoF $\downarrow$}   & Expected & \textbf{ChromaSim$\uparrow$} & Expected & \textbf{ChromaDTWS$\uparrow$} & Expected & \textbf{StructPairF$\uparrow$} & Expected & \textbf{ARI$\uparrow$}   & Expected \\
    \midrule
+7 semitones (+Perfect Fifth)
& 0.18 $\pm$ 0.09 & $\approx$ 0.167 
& 0.94 $\pm$ 0.05 & $<1$ 
& 0.81 $\pm$ 0.12 & $<1$ 
& 0.67 $\pm$ 0.14 & $\approx 1$ 
& 0.30 $\pm$ 0.29 & $\approx 1$ \\

-3 semitones (- Minor Third) 
& 0.50 $\pm$ 0.11 & 0.5--0.67 
& 0.82 $\pm$ 0.11 & $<1$ 
& 0.71 $\pm$ 0.14 & $<1$ 
& 0.69 $\pm$ 0.16 & $\approx 1$ 
& 0.37 $\pm$ 0.28 & $\approx 1$ \\

ABC$\rightarrow$AAA 
& 0.06 $\pm$ 0.13 & $\approx 0$ 
& 0.99 $\pm$ 0.01 & $\approx 1$ 
& 0.89 $\pm$ 0.12 & $\approx 1$ 
& 0.58 $\pm$ 0.17 & $<1$ 
& 0.19 $\pm$ 0.24 & $<1$ \\

ABC$\rightarrow$ABA 
& 0.05 $\pm$ 0.12 & $\approx 0$ 
& 0.99 $\pm$ 0.00 & $\approx 1$ 
& 0.92 $\pm$ 0.12 & $\approx 1$ 
& 0.65 $\pm$ 0.16 & $<1$ 
& 0.31 $\pm$ 0.27 & $<1$ \\

Tempo shifted +50\% 
& 0.03 $\pm$ 0.12 & $\approx 0$ 
& 0.99 $\pm$ 0.00 & $\approx 1$ 
& 0.84 $\pm$ 0.14 & $\approx 1$ 
& 0.66 $\pm$ 0.16 & $\approx 1$ 
& 0.25 $\pm$ 0.29 & $\approx 1$ \\

All beats shifted later by a constant -150 ms 
& 0.02 $\pm$ 0.10 & $\approx 0$ 
& 0.99 $\pm$ 0.00 & $\approx 1$ 
& 0.96 $\pm$ 0.13 & $\approx 1$ 
& 0.74 $\pm$ 0.16 & $\approx 1$ 
& 0.48 $\pm$ 0.30 & $\approx 1$ \\

Interval direction flipped for the main melody 
& 0.06 $\pm$ 0.16 & 0--0.167 
& 0.99 $\pm$ 0.02 & $<1$ 
& 0.92 $\pm$ 0.14 & $<1$ 
& 0.76 $\pm$ 0.19 & $\approx 1$ 
& 0.52 $\pm$ 0.34 & $\approx 1$ \\

The climax phrase of the main melody is inverted 
& 0.02 $\pm$ 0.11 & $\approx 0$ 
& 0.99 $\pm$ 0.00 & $\approx 1$ 
& 0.96 $\pm$ 0.13 & $\approx 1$ 
& 0.78 $\pm$ 0.17 & $\approx 1$ 
& 0.56 $\pm$ 0.32 & $\approx 1$ \\

    \midrule
    \midrule
    \textbf{Music Facets} & \multicolumn{6}{c}{\textbf{Rhythm \& Meter}}           & \multicolumn{4}{c}{\textbf{Melodic Content \& Motif}} \\
    \midrule
    \textbf{Editing} & \textbf{$\Delta$ BPM$\downarrow$} & Expected & \textbf{BeatF$\uparrow$} & Expected & \textbf{IG$\uparrow$}    & Expected & \textbf{ContourDTWS$\uparrow$} & Expected & \textbf{MotifRec$\uparrow$} & Expected \\
    \midrule
    
    +7 semitones (+Perfect Fifth) 
& 1.26 $\pm$ 8.79 & $\approx 0$ 
& 0.90 $\pm$ 0.19 & $\approx 1$ 
& 0.76 $\pm$ 0.19 & $\approx 1$ 
& 0.78 $\pm$ 0.05 & $<1$ 
& 0.61 $\pm$ 0.10 & $<1$ \\

-3 semitones (- Minor Third) 
& 0.00 $\pm$ 0.00 & $\approx 0$ 
& 0.92 $\pm$ 0.16 & $\approx 1$ 
& 0.79 $\pm$ 0.18 & $\approx 1$  
& 0.80 $\pm$ 0.06 & $<1$ 
& 0.58 $\pm$ 0.09 & $<1$ \\

ABC$\rightarrow$AAA 
& 0.87 $\pm$ 2.72 & $\approx 0$ 
& 0.54 $\pm$ 0.22 & $<1$ 
& 0.39 $\pm$ 0.17 & $<1$
& 0.76 $\pm$ 0.04 & $<1$ 
& 0.65 $\pm$ 0.09 & $<1$ \\

ABC$\rightarrow$ABA 
& 0.29 $\pm$ 1.15 & $\approx 0$ 
& 0.74 $\pm$ 0.14 & $<1$ 
& 0.57 $\pm$ 0.14 & $<1$ 
& 0.85 $\pm$ 0.05 & $<1$ 
& 0.74 $\pm$ 0.09 & $<1$ \\

Tempo shifted +50\% 
& 26.10 $\pm$ 12.96 & $>0$ 
& 0.24 $\pm$ 0.08 & $<1$ 
& 0.19 $\pm$ 0.11 & $\approx 0$ 
& 0.69 $\pm$ 0.04 & $<1$ 
& 0.60 $\pm$ 0.07 & $<1$ \\

All beats shifted later by a constant -150 ms 
& 1.35 $\pm$ 8.80 & $>0$ 
& 0.03 $\pm$ 0.07 & $\approx 0$ 
& 0.63 $\pm$ 0.13 & $<1$ 
& 0.82 $\pm$ 0.05 & $<1$ 
& 0.66 $\pm$ 0.09 & $<1$ \\

Interval direction flipped for the main melody 
& 0.13 $\pm$ 0.89 & $\approx 0$ 
& 0.94 $\pm$ 0.15 & $\approx 1$ 
& 0.83 $\pm$ 0.20 & $\approx 1$ 
& 0.78 $\pm$ 0.20 & $<1$ 
& 0.61 $\pm$ 0.30  & $<1$ \\

The climax phrase of the main melody is inverted 
& 1.26 $\pm$ 8.79 & $\approx 0$ 
& 0.98 $\pm$ 0.08 & $\approx 1$ 
& 0.93 $\pm$ 0.12 & $\approx 1$ 
& 0.95 $\pm$ 0.10 & $<1$ 
& 0.91 $\pm$ 0.17 & $<1$ \\
    \bottomrule
    \end{tabular}
    }
    \caption{Objective validation for MuseCPEval metrics.}
  \label{tab:objective_validation}%
\end{table*}%

\textbf{Motif 3-gram Recall.}\quad 
To assess whether a motif survives, we tokenize each track into a sequence of pitch-class intervals between successive notes and measure how many short interval patterns from $x$ reappear in $x'$ \cite{DBLP:conf/dlfm/FrielerBHCPD19}.
Let $\mathcal{G}_n(x)$ denote the set of all contiguous length-$n$ subsequences of this interval sequence. Each one is a short interval pattern that represents a local melodic figure. Using $n = 3$ to capture short but distinctive figures, the Motif $n$-gram Recall is defined as the fraction of reference motifs recovered in the edited output:
\begin{equation}
    \text{MotifRecall}(x, x') = \frac{|\mathcal{G}_n(x) \cap \mathcal{G}_n(x')|}{|\mathcal{G}_n(x)|}
\end{equation}
It lies in $[0, 1]$, where 1 means every motif is reproduced in $x'$, and 0 means no motif survives. 


\section{Metric Validation}
\label{sec:validation}
The metrics proposed in Section \ref{sec:MuseCP} for MuseCP evaluation are implemented by the algorithms based on probability, making it essential to verify their validity.
In this section, we conduct validation from two complementary perspectives: $i)$ objective validation using edits with known effects on specific musical facets; $ii)$ human listening studies to assess alignment with perception.

\subsection{Objective Validation}
\label{sec:objectivevalidation}
We evaluate whether the proposed metrics respond correctly to edits with known effects on specific musical facets. 
50 MIDI samples of high musicality with clear melodic lines and structure are selected from the Lakh dataset\footnote{\url{https://colinraffel.com/projects/lmd/}} for correct editing.
We apply eight tailored editing operations covering four music facets to each sample as shown in Table \ref{tab:objective_validation}.
After editing, the samples are transferred to audio.
We compute all metrics and report the mean and standard deviation across all samples in Table \ref{tab:objective_validation}, where the \textit{Expected} columns specify the anticipated metric values under each editing operation. Details are in Appendix \ref{app:objective}.

According to the results, most metrics behave consistently with expectations, demonstrating the effectiveness of our evaluation framework. 
The metrics are illustrated to be sensitive to changes in musical attributes.
For example, pitch-shifting leads to clear changes in harmonic metrics such as CoF and ChromaSim, while leaving rhythm metrics (e.g., $\Delta$BPM, BeatF) unaffected.
Conversely, tempo scaling (+50\%) results in a large $\Delta$BPM and small BeatF and IG, demonstrating strong sensitivity to rhythmic changes. 
For the structural metrics, although StructPairF and ARI show values below 1.0 for editing without effects on musical structure, such as harmony transposition and tempo shift, due to the segmentation pipeline being sensitive to spectral changes beyond structural boundaries, both metrics successfully differentiate structural edits from structure-preserving edits, confirming their utility as relative indicators.
Overall, these results show that the MuseCPEval metrics can effectively measure MuseCP capability by capturing nuanced changes in attributes.

\subsection{Human Listening Study}
\label{sec:human_listening}
\begin{table}[ht]
\centering
\scalebox{0.75}{
    \begin{tabular}{lccc}
        \toprule
        \textbf{Music Facet} & \textbf{Human--Gold} & \textbf{Metric--Human} & \textbf{Metric--Gold} \\
        \midrule
        Harmony & 72.7 & 65.9 & 93.2 \\
        Rhythm \& Meter & 100.0 & 100.0 & 100.0 \\
        Structure & 72.7 & 65.9 & 72.7 \\
        Melody \& Motif & 72.7 & 45.5 & 72.7 \\
        \bottomrule
    \end{tabular}}
\caption{Agreement rates (\%) between human perception, metric values, and gold labels across different music facets.}
\label{tab:agreement}
\end{table}
We also conduct a human listening study, as objective validation cannot guarantee that metrics align with human perception and may be biased by the limited edits we consider.
For each music facet, we design 4 pairwise comparison questions. Each question consists of one original audio clip and two edited versions with different editing strengths, which serve as gold labels indicating the relative degree of deviation from the original. 
Participants are asked to determine which edited sample is farther from the original with respect to the specified facet, while ignoring other aspects. Their choices are human labels.
The original music data used in the questionnaire are from the MIDI samples used in \S \ref{sec:objectivevalidation}.
Corresponding metric values are also computed for all origin-edited sample pairs, enabling direct comparison between metric values, gold labels, and human judgments.
We obtained 11 valid responses from human participants that passed the sanity checks.
The details are shown in the Appendix \ref{app:human}.
Finally, we report the agreement rates among human labels, gold labels, and metric-based comparisons in Table \ref{tab:agreement}.

Overall, the results demonstrate that the proposed metrics are effective and largely consistent with both gold labels and human perception. 
In particular, the Metric–Gold agreement is high across all facets, indicating that the metrics correctly capture the intended nuanced differences in music attributes. 
The Metric–Human agreement is also reasonably aligned for most facets, showing that the metrics reflect perceptually meaningful differences. 
The relatively lower agreement on melody \& motif suggests this facet is more subjective and harder to quantify. Overall, these results validate the reliability and perceptual relevance of MuseCPEval.

\section{Case Studies: Music Editing System Analysis through MuseCP}
\label{sec:setups}

While the evaluations in existing music editing works typically report aggregate quality scores, they provide limited insight into how their methodologies affect MuseCP capability and their pros and cons.
In this section, we demonstrate that our MuseCPEval metrics go beyond evaluation and serve as a diagnostic tool for music editing systems. 
We apply our framework to four representative and diverse music editing systems spanning different architectures, editing paradigms, and tasks: MusicMagus \cite{zhang2024musicmaguszeroshottexttomusicediting}, ZETA \cite{manor2024zeroshotunsupervisedtextbasedaudio}, Audio Prompt Adapter \cite{audiopromptadapter}, and Instruct-MusicGen \cite{instructmusicgen}. 
Our analysis reveals the strengths and limitations of each system through MuseCPEval, providing fine-grained insights that conventional evaluations miss.

\subsection{Setups}

\textbf{MusicMagus.}
MusicMagus is a zero-shot diffusion-based music editing method that performs text-guided latent embedding steering with cross-attention consistency constraints.
We follow its official protocol by generating 60 samples with AudioLDM2 \cite{audioldm2} and then constructing paired editing instructions through its captioning process, covering instrument and style transfer. The 60 paired editing instructions and audio are fed into MusicMagus\footnote{\url{https://github.com/ldzhangyx/MusicMagus}} to obtain edited outputs.

\textbf{ZETA.}
ZETA is a zero-shot audio editing method that inverts the input music into the latent trajectory of a pretrained diffusion model, then performs reverse denoising conditioned on a target text prompt to generate the edited output while preserving the original coarse structure.
Following its supervised text-guided editing evaluation, we use 34 musical excerpts from the MedleyDB \cite{medleydb} with 324 editing instructions. And then we directly use ZETA\footnote{\url{https://github.com/HilaManor/AudioEditingCode/}} to generate 324 edited outputs for MuseCP evaluation.

\textbf{Audio Prompt Adapter.}\quad
AP-Adapter is a lightweight adapter-based extension of AudioLDM2, which injects AudioMAE-extracted features from the input audio through decoupled cross-attention layers, enabling text-guided music editing.
Following its official evaluation\footnote{\url{https://github.com/fundwotsai2001/AP-adapter}}, we use 40 in-domain samples with 3 instrument editing instructions for each sample to conduct timbre transfer.

\textbf{Instruct-MusicGen.} \quad
Instruct-MusicGen is a text-based music editing system that extends MusicGen \cite{copet2024simplecontrollablemusicgeneration} to follow natural language instructions for modifying specific musical attributes, such as instrumentation.
Following the official setups\footnote{\url{https://github.com/ldzhangyx/instruct-MusicGen}}, we adopt instrument-level editing tasks and data from Slakh \cite{slakh}, including adding, removing, and extracting instruments. 
For each edit, 150 input audio and editing instruction pairs are evaluated.

    

\begin{figure*}
    \centering
    \includegraphics[width=0.84\linewidth]{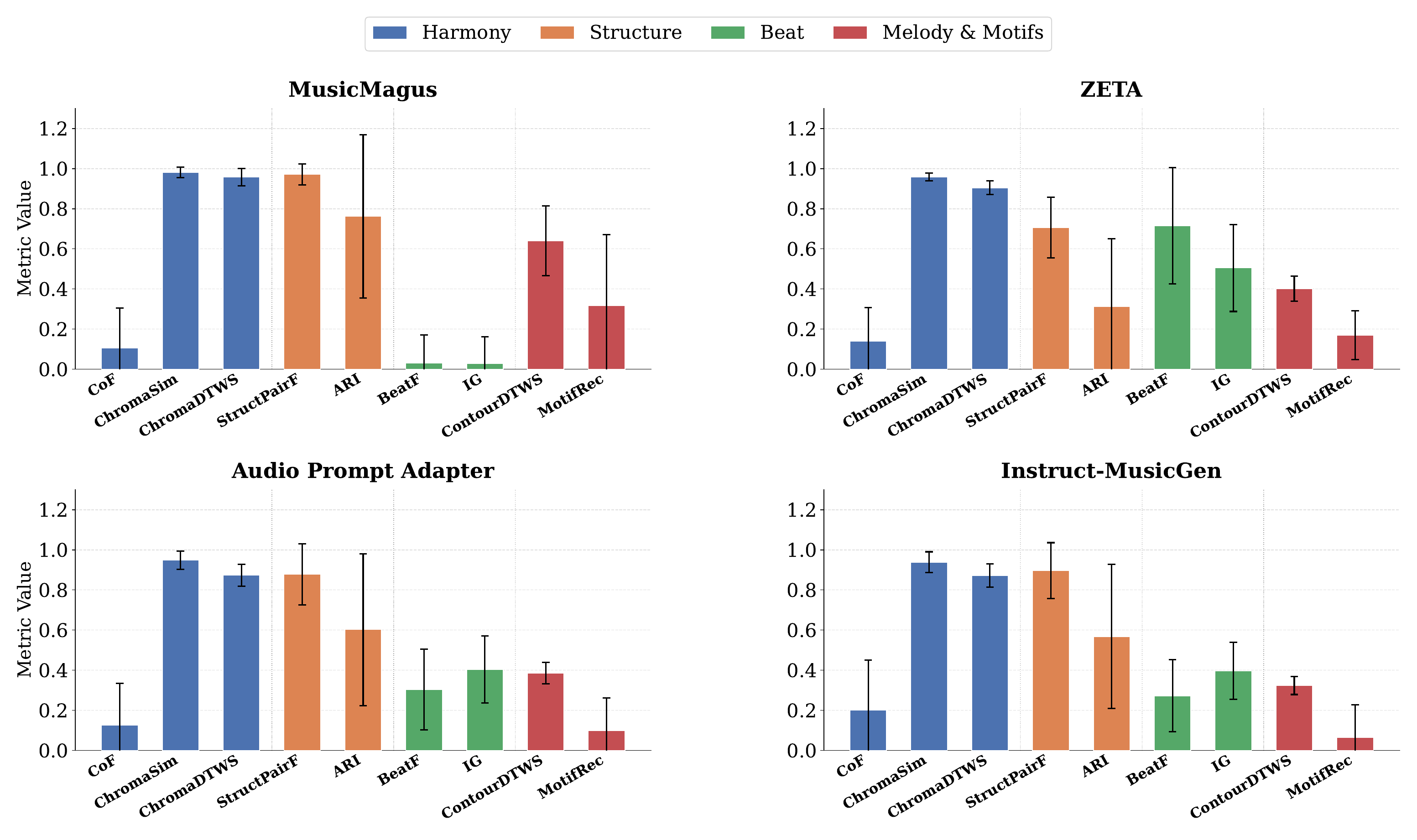}
    \vspace{-3mm}
    \caption{MuseCPEval evaluation from four different music editing systems. Note that these systems cannot be compared because the evaluation setups of the systems are different from each other.}
    \label{fig:eot1}
\end{figure*}

\subsection{Case Analysis}

\textbf{MusicMagus.}\quad
MusicMagus demonstrates strong MuseCP ability in harmony and structural form, achieving near-perfect chroma-based similarity, such as ChromaSim and ChromaDTWS, and high structural consistency (StructPairF).
These results suggest that the model effectively retains global harmonic content and large-scale arrangement while modifying target attributes. 
It also shows relatively good preservation of melodic contour and moderate motif retention according to ContourDTWS and MotifRec, respectively.
They suggest that enforcing cross-attention consistency during diffusion sampling in MusicMagus significantly contributes to MuseCP. 
The low $\Delta$BPM shows MusicMagus performs well on rhythm preservation, while its BeatF and IG are near-zero, which we think may reflect musically meaningful rhythmic transformations induced by style-transfer edits rather than unintended context loss.
Meanwhile, the strong overall MuseCP capability suggests that MusicMagus' inference-time intervention, without additional fine-tuning, could avoid distorting the original musical content.

\begin{table}[ht]
\centering
\scalebox{0.75}{
\begin{tabular}{lc}
\toprule
\textbf{Music Editing System} & \textbf{$\Delta$ BPM} \\ 
\midrule
MusicMagus & $5.573 \pm 10.315$ \\
ZETA & $4.253 \pm 12.234$ \\
Audio Prompt Adapter & $20.856 \pm 20.584$ \\
Instruct-MusicGen & $20.012 \pm 14.310$ \\
\bottomrule
\end{tabular}}
\caption{$\Delta$BPM for each music editing systems.}
\label{tab:delta_bpm}
\end{table}

\textbf{ZETA.}\quad
ZETA adopts a zero-shot DDPM inversion framework, where the edited audio is generated by reusing the inverted latent trajectory of the source signal while replacing the conditioning text prompt. This design naturally encourages preservation of low- and mid-level musical context inherited from the original audio. 
Consistent with this mechanism, ZETA demonstrates particularly strong harmony and beat preservation in our MuseCP evaluation. 
High chroma-based similarity scores suggest that the tonal center and harmonic content are largely retained after editing, while strong beat-related metrics indicate that temporal periodicity, onset patterns, and rhythmic pulse remain stable throughout the generation process. 
These results imply that the inversion process effectively anchors the edited output to the source music, allowing semantic modifications without substantially disrupting core harmonic or rhythmic foundations. 
At the same time, ZETA’s prompt-driven resampling may still alter higher-level musical organization or melodic details when stronger edits are applied, highlighting a common trade-off in text-guided diffusion editing between editability and preservation of fine-grained long-range context. 
Overall, the MuseCP results suggest that ZETA is especially effective when the desired edit should preserve the original harmonic progression and rhythmic groove.

\textbf{Audio Prompt Adapter}.\quad
This system demonstrates a preservation profile that strongly aligns with its adapter-based conditioning design. 
By injecting source-audio representations into a pretrained text-to-music diffusion backbone through decoupled cross-attention, it explicitly encourages reuse of high-level contextual information from the input audio. 
This is reflected in its strong harmony and structure preservation, achieving competitive scores on 0.95 ChromaSim, 0.88 StructPairF, and the relatively high ARI, suggesting that harmonic content and large-scale form are largely retained during editing. 
However, its beat and melody preservation are notably weaker, with very large tempo deviations, high $\Delta$BPM, and only moderate ContourDTW.
We suggest that global semantic audio embeddings provide weaker constraints on fine-grained temporal alignment and exact melodic trajectories during diffusion sampling. 
Overall, Audio Prompt Adapter represents a strong solution for high-level context preservation, while highlighting the need for future editing systems to better model local temporal and melodic consistency.

\textbf{Instruct-MusicGen.}\quad
Instruct-MusicGen shows a facet-dependent preservation pattern under instrument-level editing. 
It preserves harmony and structure relatively well, but shows clear degradation in rhythm \& meter and melodic content, which is consistent with its parameter-efficient architecture. 
Instruct-MusicGen extends MusicGen with an audio fusion module and a text fusion module, while keeping most pretrained components frozen and only training lightweight fusion parameters and LoRA modules in the text cross-attention. 
Such a design is effective for injecting the source audio and instruction into a strong pretrained MusicGen backbone, which helps retain coarse musical context such as global tonal distribution and segment-level organization. 
However, because the model is not explicitly constrained to align beat positions, tempo, or melodic trajectories with the input, the autoregressive generation process can still introduce timing drift and local melodic changes, leading to poor $\Delta$BPM, BeatF, and IG,  and lower ContourDTW/MotifRec scores. 
The melody degradation should also be interpreted together with the task setting.
In adding, removing, and extracting instrument editing, the target stem may carry part of the main melody, so a semantically correct edit can naturally alter melodic contours and motif statistics. 
Overall, these results suggest that Instruct-MusicGen is effective for instruction-following stem manipulation and coarse context retention, but its lightweight fusion strategy is less sufficient for fine-grained MuseCP, especially for time-locked rhythmic structure and melody preservation.

\section{Conclusion}
In this work, we introduced MuseCPEval, the 1st music context preservation evaluation benchmark to measure the capability of music editing systems to preserve music attributes that are not intended to be changed during editing.
10 metrics covering four music facets are collected and tailored for comprehensive evaluation.
Through objective validation and a human listening study, it is shown that these metrics are effective at capturing fine-grained changes in music attributes for MuseCP evaluation.
The case studies on four representative music editing systems with MuseCPEval offer practical insights.
We hope our framework can serve as a testbed to understand the cons and pros of music editing systems and provide insights for constructing more reliable editing systems in the future.

\newpage

\section{AI Usage and Ethics Statement}
ChatGPT is used only for limited language polishing.
All technical content, experimental design, implementation, result organization, and conclusions were developed by the human authors.
This work presents evaluation metrics for music editing systems and does not involve high-risk applications. 
For the human listening study, participation was voluntary and anonymous, and no sensitive personal data was collected. 
Participants were fairly compensated. 
All data and code used are publicly available.

\bibliographystyle{IEEEtran}
\bibliography{ISMIRtemplate}

@inproceedings{DBLP:conf/ismir/BittnerSEB15,
  author       = {Rachel M. Bittner and
                  Justin Salamon and
                  Slim Essid and
                  Juan Pablo Bello},
  editor       = {Meinard M{\"{u}}ller and
                  Frans Wiering},
  title        = {Melody Extraction by Contour Classification},
  booktitle    = {Proceedings of the 16th International Society for Music Information
                  Retrieval Conference, {ISMIR} 2015, M{\'{a}}laga, Spain, October
                  26-30, 2015},
  pages        = {500--506},
  year         = {2015},
  url          = {http://ismir2015.uma.es/articles/247\_Paper.pdf},
  bibsource    = {dblp computer science bibliography, https://dblp.org}
}

@inproceedings{DBLP:conf/dlfm/FrielerBHCPD19,
  author       = {Klaus Frieler and
                  Dogac Basaran and
                  Frank H{\"{o}}ger and
                  H{\'{e}}l{\`{e}}ne{-}Camille Crayencour and
                  Geoffroy Peeters and
                  Simon Dixon},
  editor       = {David Rizo},
  title        = {Don't hide in the frames: Note- and pattern-based evaluation of automated
                  melody extraction algorithms},
  booktitle    = {6th International Conference on Digital Libraries for Musicology,
                  DLfM 2019, The Hague, The Netherlands, November 19, 2019},
  pages        = {25--32},
  publisher    = {{ACM}},
  year         = {2019},
  url          = {https://doi.org/10.1145/3358664.3358672},
  doi          = {10.1145/3358664.3358672},
  bibsource    = {dblp computer science bibliography, https://dblp.org}
}

@article{6fd45ff5-8145-3f39-a258-8e3ef378c6a4,
 ISSN = {01621459, 1537274X},
 URL = {http://www.jstor.org/stable/2284239},
 author = {William M. Rand},
 journal = {Journal of the American Statistical Association},
 number = {336},
 pages = {846--850},
 publisher = {[American Statistical Association, Taylor & Francis, Ltd.]},
 title = {Objective Criteria for the Evaluation of Clustering Methods},
 urldate = {2025-09-15},
 volume = {66},
 year = {1971}
}

@inproceedings{Raffel2014MIR_EVALAT,
  author       = {Colin Raffel and
                  Brian McFee and
                  Eric J. Humphrey and
                  Justin Salamon and
                  Oriol Nieto and
                  Dawen Liang and
                  Daniel P. W. Ellis},
  editor       = {Hsin{-}Min Wang and
                  Yi{-}Hsuan Yang and
                  Jin Ha Lee},
  title        = {MIR{\_}EVAL: {A} Transparent Implementation of Common {MIR} Metrics},
  booktitle    = {Proceedings of the 15th International Society for Music Information
                  Retrieval Conference, {ISMIR} 2014, Taipei, Taiwan, October 27-31,
                  2014},
  pages        = {367--372},
  year         = {2014},
  url          = {http://www.terasoft.com.tw/conf/ismir2014/proceedings/T066\_320\_Paper.pdf},
  bibsource    = {dblp computer science bibliography, https://dblp.org}
}

@article{DBLP:journals/tmm/ZhuK06,
  author       = {Yongwei Zhu and
                  Mohan S. Kankanhalli},
  title        = {Precise pitch profile feature extraction from musical audio for key
                  detection},
  journal      = {{IEEE} Trans. Multim.},
  volume       = {8},
  number       = {3},
  pages        = {575--584},
  year         = {2006},
  url          = {https://doi.org/10.1109/TMM.2006.870727},
  doi          = {10.1109/TMM.2006.870727},
  bibsource    = {dblp computer science bibliography, https://dblp.org}
}

@misc{kader2025surveyevaluationmetricsmusic,
  author       = {Faria Binte Kader and
                  Santu Karmaker},
  title        = {A Survey on Evaluation Metrics for Music Generation},
  journal      = {CoRR},
  volume       = {abs/2509.00051},
  year         = {2025},
  url          = {https://doi.org/10.48550/arXiv.2509.00051},
  doi          = {10.48550/ARXIV.2509.00051},
  eprinttype   = {arXiv},
  eprint       = {2509.00051},
  bibsource    = {dblp computer science bibliography, https://dblp.org}
}

@inproceedings{copet2024simplecontrollablemusicgeneration,
title={Simple and Controllable Music Generation},
author={Jade Copet and Felix Kreuk and Itai Gat and Tal Remez and David Kant and Gabriel Synnaeve and Yossi Adi and Alexandre D{\'e}fossez},
booktitle={Thirty-seventh Conference on Neural Information Processing Systems},
year={2023},
url={https://openreview.net/forum?id=jtiQ26sCJi}
}

@inproceedings{
wang2023audit,
title={{AUDIT}: Audio Editing by Following Instructions with Latent Diffusion Models},
author={Yuancheng Wang and Zeqian Ju and Xu Tan and Lei He and Zhizheng Wu and Jiang Bian and sheng zhao},
booktitle={Thirty-seventh Conference on Neural Information Processing Systems},
year={2023},
url={https://openreview.net/forum?id=EO1KuHoR0V}
}

@inproceedings{instructme,
  author       = {Bing Han and
                  Junyu Dai and
                  Weituo Hao and
                  Xinyan He and
                  Dong Guo and
                  Jitong Chen and
                  Yuxuan Wang and
                  Yanmin Qian and
                  Xuchen Song},
  title        = {InstructME: An Instruction Guided Music Edit Framework with Latent
                  Diffusion Models},
  booktitle    = {Proceedings of the Thirty-Third International Joint Conference on
                  Artificial Intelligence, {IJCAI} 2024, Jeju, South Korea, August 3-9,
                  2024},
  pages        = {5835--5843},
  publisher    = {ijcai.org},
  year         = {2024},
  url          = {https://www.ijcai.org/proceedings/2024/645},
  bibsource    = {dblp computer science bibliography, https://dblp.org}
}

@inproceedings{zhang2024musicmaguszeroshottexttomusicediting,
  author       = {Yixiao Zhang and
                  Yukara Ikemiya and
                  Gus Xia and
                  Naoki Murata and
                  Marco A. Mart{\'{\i}}nez Ram{\'{\i}}rez and
                  Wei{-}Hsiang Liao and
                  Yuki Mitsufuji and
                  Simon Dixon},
  title        = {MusicMagus: Zero-Shot Text-to-Music Editing via Diffusion Models},
  booktitle    = {Proceedings of the Thirty-Third International Joint Conference on
                  Artificial Intelligence, {IJCAI} 2024, Jeju, South Korea, August 3-9,
                  2024},
  pages        = {7805--7813},
  publisher    = {ijcai.org},
  year         = {2024},
  url          = {https://www.ijcai.org/proceedings/2024/864},
  bibsource    = {dblp computer science bibliography, https://dblp.org}
}

@inproceedings{ramoneda2024musicproofreadingrefinpaintmodify,
  author       = {Pedro Ramoneda and
                  Mart{\'{\i}}n Rocamora and
                  Taketo Akama},
  editor       = {Blair Kaneshiro and
                  Gautham J. Mysore and
                  Oriol Nieto and
                  Chris Donahue and
                  Cheng{-}Zhi Anna Huang and
                  Jin Ha Lee and
                  Brian McFee and
                  Matthew C. McCallum},
  title        = {Music Proofreading With RefinPaint: Where and How to Modify Compositions
                  Given Context},
  booktitle    = {Proceedings of the 25th International Society for Music Information
                  Retrieval Conference, {ISMIR} 2024, San Francisco, California, {USA}
                  and Online, November 10-14, 2024},
  pages        = {240--247},
  year         = {2024},
  url          = {https://doi.org/10.5281/zenodo.14877319},
  doi          = {10.5281/ZENODO.14877319},
  bibsource    = {dblp computer science bibliography, https://dblp.org}
}

@inproceedings{manor2024zeroshotunsupervisedtextbasedaudio,
  author       = {Hila Manor and
                  Tomer Michaeli},
  editor       = {Ruslan Salakhutdinov and
                  Zico Kolter and
                  Katherine A. Heller and
                  Adrian Weller and
                  Nuria Oliver and
                  Jonathan Scarlett and
                  Felix Berkenkamp},
  title        = {Zero-Shot Unsupervised and Text-Based Audio Editing Using {DDPM} Inversion},
  booktitle    = {Forty-first International Conference on Machine Learning, {ICML} 2024,
                  Vienna, Austria, July 21-27, 2024},
  series       = {Proceedings of Machine Learning Research},
  pages        = {34603--34629},
  publisher    = {{PMLR} / OpenReview.net},
  year         = {2024},
  url          = {https://proceedings.mlr.press/v235/manor24a.html},
  bibsource    = {dblp computer science bibliography, https://dblp.org}
}

@inproceedings{steermusic,
  author       = {Xinlei Niu and
                  Kin Wai Cheuk and
                  Jing Zhang and
                  Naoki Murata and
                  Chieh{-}Hsin Lai and
                  Michele Mancusi and
                  Woosung Choi and
                  Giorgio Fabbro and
                  Wei{-}Hsiang Liao and
                  Charles Patrick Martin and
                  Yuki Mitsufuji},
  editor       = {Sven Koenig and
                  Chad Jenkins and
                  Matthew E. Taylor},
  title        = {SteerMusic: Enhanced Musical Consistency for Zero-shot Text-Guided
                  and Personalized Music Editing},
  booktitle    = {Fortieth {AAAI} Conference on Artificial Intelligence, Thirty-Eighth
                  Conference on Innovative Applications of Artificial Intelligence,
                  Sixteenth Symposium on Educational Advances in Artificial Intelligence,
                  {AAAI} 2026, Singapore, January 20-27, 2026},
  pages        = {2000--2010},
  publisher    = {{AAAI} Press},
  year         = {2026},
  url          = {https://doi.org/10.1609/aaai.v40i3.37181},
  doi          = {10.1609/AAAI.V40I3.37181},
  bibsource    = {dblp computer science bibliography, https://dblp.org}
}

@inproceedings{instructmusicgen,
  author       = {Yixiao Zhang and
                  Yukara Ikemiya and
                  Woosung Choi and
                  Naoki Murata and
                  Marco A. Mart{\'{\i}}nez Ram{\'{\i}}rez and
                  Liwei Lin and
                  Gus Xia and
                  Wei{-}Hsiang Liao and
                  Yuki Mitsufuji and
                  Simon Dixon},
  editor       = {Juhan Nam and
                  Dasaem Jeong and
                  Keunwoo Choi and
                  Li Su and
                  Magdalena Fuentes and
                  Tomoyasu Nakano and
                  Xiao Hu and
                  Hao{-}Wen (Herman) Dong},
  title        = {Instruct-MusicGen: Unlocking Text-to-Music Editing for Music Language
                  Models via Instruction Tuning},
  booktitle    = {Proceedings of the 26th International Society for Music Information
                  Retrieval Conference, {ISMIR} 2025, Daejeon, South Korea, September
                  21-25, 2025},
  pages        = {328--336},
  year         = {2025},
  url          = {https://doi.org/10.5281/zenodo.17811375},
  doi          = {10.5281/ZENODO.17811375},
  bibsource    = {dblp computer science bibliography, https://dblp.org}
}

@article{notthatgroove,
  author       = {Li Zhang},
  title        = {Not that Groove: Zero-Shot Symbolic Music Editing},
  journal      = {CoRR},
  volume       = {abs/2505.08203},
  year         = {2025},
  url          = {https://doi.org/10.48550/arXiv.2505.08203},
  doi          = {10.48550/ARXIV.2505.08203},
  eprinttype   = {arXiv},
  eprint       = {2505.08203},
  bibsource    = {dblp computer science bibliography, https://dblp.org}
}

@inproceedings{medleydb,
  author       = {Rachel M. Bittner and
                  Justin Salamon and
                  Mike Tierney and
                  Matthias Mauch and
                  Chris Cannam and
                  Juan Pablo Bello},
  title        = {MedleyDB: {A} Multitrack Dataset for Annotation-Intensive {MIR} Research},
  booktitle    = {{ISMIR}},
  pages        = {155--160},
  year         = {2014}
}

@book{10.1093/acprof:oso/9780195148367.001.0001,
    author = {Krumhansl, Carol L.},
    title = {Cognitive Foundations of Musical Pitch},
    publisher = {Oxford University Press},
    year = {2001},
    month = {11},
    isbn = {9780195148367},
    doi = {10.1093/acprof:oso/9780195148367.001.0001},
    url = {https://doi.org/10.1093/acprof:oso/9780195148367.001.0001},
}

@article{Brown1991CalculationOA,
  title={Calculation of a constant Q spectral transform},
  author={Judith C. Brown},
  journal={Journal of the Acoustical Society of America},
  year={1991},
  volume={89},
  pages={425-434},
  url={https://api.semanticscholar.org/CorpusID:44239681}
}

@inbook{dtw,
author = {Sakoe, Hiroaki and Chiba, Seibi},
title = {Dynamic programming algorithm optimization for spoken word recognition},
year = {1990},
isbn = {1558601244},
publisher = {Morgan Kaufmann Publishers Inc.},
address = {San Francisco, CA, USA},
booktitle = {Readings in Speech Recognition},
pages = {159–165},
numpages = {7}
}

@inproceedings{slakh,
  author       = {Ethan Manilow and
                  Gordon Wichern and
                  Prem Seetharaman and
                  Jonathan Le Roux},
  title        = {Cutting Music Source Separation Some Slakh: {A} Dataset to Study the
                  Impact of Training Data Quality and Quantity},
  booktitle    = {2019 {IEEE} Workshop on Applications of Signal Processing to Audio
                  and Acoustics, {WASPAA} 2019, New Paltz, NY, USA, October 20-23, 2019},
  pages        = {45--49},
  publisher    = {{IEEE}},
  year         = {2019},
  url          = {https://doi.org/10.1109/WASPAA.2019.8937170},
  doi          = {10.1109/WASPAA.2019.8937170},
  bibsource    = {dblp computer science bibliography, https://dblp.org}
}

@inproceedings{audiopromptadapter,
  author       = {Fang{-}Duo Tsai and
                  Shih{-}Lun Wu and
                  Haven Kim and
                  Bo{-}Yu Chen and
                  Hao{-}Chung Cheng and
                  Yi{-}Hsuan Yang},
  editor       = {Blair Kaneshiro and
                  Gautham J. Mysore and
                  Oriol Nieto and
                  Chris Donahue and
                  Cheng{-}Zhi Anna Huang and
                  Jin Ha Lee and
                  Brian McFee and
                  Matthew C. McCallum},
  title        = {Audio Prompt Adapter: Unleashing Music Editing Abilities for Text-to-Music
                  With Lightweight Finetuning},
  booktitle    = {Proceedings of the 25th International Society for Music Information
                  Retrieval Conference, {ISMIR} 2024, San Francisco, California, {USA}
                  and Online, November 10-14, 2024},
  pages        = {634--641},
  year         = {2024},
  url          = {https://doi.org/10.5281/zenodo.14877417},
  doi          = {10.5281/ZENODO.14877417},
  bibsource    = {dblp computer science bibliography, https://dblp.org}
}

@inproceedings{melodia,
  author       = {Yi Yang and
                  Haowen Li and
                  Tianxiang Li and
                  Boyu Cao and
                  Xiaohan Zhang and
                  Liqun Chen and
                  Qi Liu},
  editor       = {Sven Koenig and
                  Chad Jenkins and
                  Matthew E. Taylor},
  title        = {Melodia: Training-Free Music Editing Guided by Attention Probing in
                  Diffusion Models},
  booktitle    = {Fortieth {AAAI} Conference on Artificial Intelligence, Thirty-Eighth
                  Conference on Innovative Applications of Artificial Intelligence,
                  Sixteenth Symposium on Educational Advances in Artificial Intelligence,
                  {AAAI} 2026, Singapore, January 20-27, 2026},
  pages        = {2209--2217},
  publisher    = {{AAAI} Press},
  year         = {2026},
  url          = {https://doi.org/10.1609/aaai.v40i3.37204},
  doi          = {10.1609/AAAI.V40I3.37204},
  bibsource    = {dblp computer science bibliography, https://dblp.org}
}

@article{melodyflow,
  author       = {Ga{\"{e}}l Le Lan and
                  Bowen Shi and
                  Zhaoheng Ni and
                  Sidd Srinivasan and
                  Anurag Kumar and
                  Brian Ellis and
                  David Kant and
                  Varun Nagaraja and
                  Ernie Chang and
                  Wei{-}Ning Hsu and
                  Yangyang Shi and
                  Vikas Chandra},
  title        = {High Fidelity Text-Guided Music Editing via Single-Stage
                  Flow Matching},
  journal      = {CoRR},
  volume       = {abs/2407.03648},
  year         = {2024},
  url          = {https://doi.org/10.48550/arXiv.2407.03648},
  doi          = {10.48550/ARXIV.2407.03648},
  eprinttype   = {arXiv},
  eprint       = {2407.03648},
  bibsource    = {dblp computer science bibliography, https://dblp.org}
}

@inproceedings{editingmusicwithmelodyandtext,
  author       = {Siyuan Hou and
                  Shansong Liu and
                  Ruibin Yuan and
                  Wei Xue and
                  Ying Shan and
                  Mangsuo Zhao and
                  Chao Zhang},
  title        = {Editing Music with Melody and Text: Using ControlNet for Diffusion
                  Transformer},
  booktitle    = {2025 {IEEE} International Conference on Acoustics, Speech and Signal
                  Processing, {ICASSP} 2025, Hyderabad, India, April 6-11, 2025},
  pages        = {1--5},
  publisher    = {{IEEE}},
  year         = {2025},
  url          = {https://doi.org/10.1109/ICASSP49660.2025.10890309},
  doi          = {10.1109/ICASSP49660.2025.10890309},
  bibsource    = {dblp computer science bibliography, https://dblp.org}
}

@inproceedings{CQT,
  title={Constant-Q transform toolbox for music processing},
  author={Christian Schoerkhuber and Anssi Klapuri},
  year={2010},
  publisher={7th Sound and Music Computing Conference},
  url={https://api.semanticscholar.org/CorpusID:12358579}
}

@inproceedings{librosa,
  author       = {Brian McFee and
                  Colin Raffel and
                  Dawen Liang and
                  Daniel P. W. Ellis and
                  Matt McVicar and
                  Eric Battenberg and
                  Oriol Nieto},
  editor       = {Kathryn D. Huff and
                  James Bergstra},
  title        = {librosa: Audio and Music Signal Analysis in Python},
  booktitle    = {Proceedings of the 14th Python in Science Conference, SciPy 2015,
                  Austin, Texas, USA, July 6-12, 2015},
  pages        = {18--24},
  publisher    = {scipy.org},
  year         = {2015},
  url          = {https://doi.org/10.25080/Majora-7b98e3ed-003},
  doi          = {10.25080/MAJORA-7B98E3ED-003},
  bibsource    = {dblp computer science bibliography, https://dblp.org}
}

@inproceedings{msaf,
  author       = {Oriol Nieto and
                  Juan Pablo Bello},
  editor       = {Michael I. Mandel and
                  Johanna Devaney and
                  Douglas Turnbull and
                  George Tzanetakis},
  title        = {Systematic Exploration of Computational Music Structure Research},
  booktitle    = {Proceedings of the 17th International Society for Music Information
                  Retrieval Conference, {ISMIR} 2016, New York City, United States,
                  August 7-11, 2016},
  pages        = {547--553},
  year         = {2016},
  url          = {https://wp.nyu.edu/ismir2016/wp-content/uploads/sites/2294/2016/07/043\_Paper.pdf},
  bibsource    = {dblp computer science bibliography, https://dblp.org}
}

@article{audioldm2,
  author       = {Haohe Liu and
                  Yi Yuan and
                  Xubo Liu and
                  Xinhao Mei and
                  Qiuqiang Kong and
                  Qiao Tian and
                  Yuping Wang and
                  Wenwu Wang and
                  Yuxuan Wang and
                  Mark D. Plumbley},
  title        = {AudioLDM 2: Learning Holistic Audio Generation With Self-Supervised
                  Pretraining},
  journal      = {{IEEE} {ACM} Trans. Audio Speech Lang. Process.},
  volume       = {32},
  pages        = {2871--2883},
  year         = {2024},
  url          = {https://doi.org/10.1109/TASLP.2024.3399607},
  doi          = {10.1109/TASLP.2024.3399607},
  bibsource    = {dblp computer science bibliography, https://dblp.org}
}

@inproceedings{mfcc,
  author       = {Beth Logan},
  title        = {Mel Frequency Cepstral Coefficients for Music Modeling},
  booktitle    = {{ISMIR} 2000, 1st International Symposium on Music Information Retrieval,
                  Plymouth, Massachusetts, USA, October 23-25, 2000, Proceedings},
  year         = {2000},
  url          = {http://ismir2000.ismir.net/papers/logan\_paper.pdf},
  bibsource    = {dblp computer science bibliography, https://dblp.org}
}

@inproceedings{Mandel2005SongLevelFA,
  title={Song-Level Features and Support Vector Machines for Music Classification},
  author={Michael I. Mandel and Daniel P. W. Ellis},
  booktitle={International Society for Music Information Retrieval Conference},
  year={2005},
  url={https://archives.ismir.net/ismir2005/paper/000106.pdf}
}

@inproceedings{mundada-etal-2025-wildscore,
    title = "{W}ild{S}core: Benchmarking {MLLM}s in-the-Wild Symbolic Music Reasoning",
    author = "Mundada, Gagan  and
      Vishe, Yash  and
      Namburi, Amit  and
      Xu, Xin  and
      Novack, Zachary  and
      McAuley, Julian  and
      Wu, Junda",
    editor = "Christodoulopoulos, Christos  and
      Chakraborty, Tanmoy  and
      Rose, Carolyn  and
      Peng, Violet",
    booktitle = "Proceedings of the 2025 Conference on Empirical Methods in Natural Language Processing",
    month = nov,
    year = "2025",
    address = "Suzhou, China",
    publisher = "Association for Computational Linguistics",
    url = "https://aclanthology.org/2025.emnlp-main.853/",
    doi = "10.18653/v1/2025.emnlp-main.853",
    pages = "16847--16863",
    ISBN = "979-8-89176-332-6"
}

@article{wang2025csymr,
  title={CSyMR: Benchmarking Compositional Music Information Retrieval in Symbolic Music Reasoning},
  author={Wang, Boyang and Vishe, Yash and Xu, Xin and Novack, Zachary and Jiang, Xunyi and McAuley, Julian and Wu, Junda},
  journal={arXiv preprint arXiv:2601.11556},
  year={2025}
}

%
%
%
%
\appendix

\begin{table*}[ht!]
\centering
\begin{tabular}{lcccc}
\toprule
 & \multicolumn{4}{c}{\textbf{Timbre}} \\
\cmidrule(lr){2-5}
\textbf{Editing} & SKLSim$\uparrow$ & Expected & MMCos$\uparrow$ & Expected \\
\midrule
Grand Piano $\rightarrow$ Electric Piano          & 0.16 $\pm$ 0.15 & $<$ 1      & 0.78 $\pm$ 0.07 & $<$ 1      \\
Grand Piano $\rightarrow$ Acoustic Guitar         & 0.35 $\pm$ 0.12 & $<$ 1      & 0.90 $\pm$ 0.04 & $<$ 1      \\
\midrule
+7 semitones (+Perfect Fifth)                     & 0.72 $\pm$ 0.13 & $\approx$ 1 & 0.97 $\pm$ 0.05 & $\approx$ 1 \\
-3 semitones (- Minor Third)                      & 0.84 $\pm$ 0.11 & $\approx$ 1 & 0.99 $\pm$ 0.02 & $\approx$ 1 \\
ABC$\rightarrow$AAA                               & 0.94 $\pm$ 0.13 & $\approx$ 1 & 1.00 $\pm$ 0.01 & $\approx$ 1 \\
ABC$\rightarrow$ABA                               & 0.97 $\pm$ 0.11 & $\approx$ 1 & 1.00 $\pm$ 0.01 & $\approx$ 1 \\
Tempo shifted +50\%                               & 0.87 $\pm$ 0.13 & $\approx$ 1 & 0.99 $\pm$ 0.02 & $\approx$ 1 \\
All beats shifted later by a constant -150ms      & 0.97 $\pm$ 0.11 & $\approx$ 1 & 1.00 $\pm$ 0.01 & $\approx$ 1 \\
Interval direction flipped for the main melody    & 0.87 $\pm$ 0.16 & $\approx$ 1 & 0.99 $\pm$ 0.02 & $\approx$ 1 \\
The climax phrase of the main melody is inverted  & 0.98 $\pm$ 0.11 & $\approx$ 1 & 1.00 $\pm$ 0.01 & $\approx$ 1 \\
\bottomrule
\end{tabular}
\caption{Objective validation for the timbre metrics, where the two instrument-replacement editings are referenced against the Grand Piano render of each sample.}
\label{tab:timbre}
\end{table*}

\section{Timbre Preservation Evaluation}
Thanks to ISMIR reviewers' suggestions, we also explore timbre context preservation. 
Timbre is the attribute by which listeners distinguish two sounds of the same pitch, loudness, and duration, and it is the primary cue for identifying which instrument is playing. 
We evaluate timbre context preservation, i.e., whether $x'$ still sounds as though it were played on the same instruments as $x$.

\textbf{SKL Similarity.}\quad
This metric compares the global timbral content of $x$ and $x'$ \cite{Mandel2005SongLevelFA}.
Concretely, $x$ and $x'$ are each converted into an MFCC \cite{mfcc} feature sequence $\mathbf{M} \in \mathbb{R}^{T \times F}$ and $\mathbf{M}' \in \mathbb{R}^{T' \times F}$, where the $0$-th cepstral coefficient is discarded because it encodes overall energy rather than spectral shape, and the remaining 12 coefficients are stacked with their first and second temporal derivatives, giving $F=36$. 
Each sequence is summarized by a single multivariate
Gaussian fitted to its frames,
$\mathcal{N}(\boldsymbol{\mu}, \boldsymbol{\Sigma})$ and
$\mathcal{N}(\boldsymbol{\mu}', \boldsymbol{\Sigma}')$, where
$\boldsymbol{\mu} \in \mathbb{R}^{F}$ and
$\boldsymbol{\Sigma} \in \mathbb{R}^{F \times F}$ are the sample mean and covariance of the rows of $\mathbf{M}$, and likewise for $\mathbf{M}'$.The timbral distance is the symmetrized KL divergence between them, which admits the closed form $\text{SKL}(x,x') = \tfrac{1}{2}\Big[\mathrm{tr}(\boldsymbol{\Sigma}'^{-1}\boldsymbol{\Sigma}) + \mathrm{tr}(\boldsymbol{\Sigma}^{-1}\boldsymbol{\Sigma}') + \boldsymbol{\delta}^{\top}(\boldsymbol{\Sigma}^{-1} + \boldsymbol{\Sigma}'^{-1})\boldsymbol{\delta} - 2F\Big]$
where $\boldsymbol{\delta} = \boldsymbol{\mu}' - \boldsymbol{\mu}$ and the log-determinant terms cancel in the symmetric sum. Since $\text{SKL}$ is unbounded and heavy-tailed, we map it to a bounded similarity:
\begin{equation}
    \text{SKLSim}(x,x') = \frac{1}{1 + \text{SKL}(x,x')/50}
\end{equation}
It lies in $[0,1]$, where $1$ indicates identical timbral distributions and values near $0$ indicate that the two recordings occupy disjoint regions of timbre space.

\textbf{Mean MFCC Cosine.}\quad
To grade the magnitude of editing, we additionally compare the average spectral envelopes of $x$ and $x'$.
Let $\bar{\mathbf{m}} = \frac{1}{T}\sum_{t=1}^{T} \mathbf{m}_t$ and $\bar{\mathbf{m}}'$ denote the time-averaged MFCC frame vectors of $x$ and $x'$. The Mean MFCC Cosine is their cosine similarity:
\begin{equation}
    \text{MMCos}(x,x') = \frac{\bar{\mathbf{m}} \cdot \bar{\mathbf{m}}'}{|\bar{\mathbf{m}}||\bar{\mathbf{m}}'|}
\end{equation}
It lies in $[-1,1]$, where $1$ means the two recordings share the same average spectral envelope. 

In addition to the editing curated in Section~\ref{sec:objectivevalidation}, we also tailor two editing for timbre context preservation evaluation: changing the instrument, the grand piano, into other instruments, the electric piano and acoustic guitar.
Table \ref{tab:timbre} shows the results.

\section{Objective Validation Setups}
\label{app:objective}
This section provides the details of the objective validation referenced in Section~\ref{sec:objectivevalidation}.

\subsection{Definition of the Editing Operations}
\label{app:objective_edits}
Table~\ref{tab:edit_definitions} gives the transformation applied by each of the eight operations abbreviated in Table~\ref{tab:objective_validation} and Table \ref{tab:timbre}, together with the facet it targets.

\subsection{Evaluation Set and Audio Rendering}
\label{app:objective_data}
The 50 MIDI files are drawn from the Lakh MIDI Dataset\footnote{\url{https://colinraffel.com/projects/lmd/}} and selected to exhibit a clearly identifiable main melodic line, a stable metrical structure, and a recognizable sectional form, so that a facet-targeted edit yields a well-defined change in the corresponding facet.
Editing in the symbolic domain is deliberate.
It lets each operation alter exactly one intended facet while leaving the others untouched by construction, which is the controlled condition required to test whether a metric responds only to the facet it is designed to measure.
The main melody used by the two melodic operations is taken to be the monophonic track with the highest mean pitch among the tracks carrying continuous note activity, and the three sections used by the two structural operations are taken from the sectional annotation of each source, so that both are defined on the symbolic score rather than recovered from audio.
Each source yields ten edited variants, giving $50 \times 10 = 500$ origin-edited pairs.
Both the original and the edited MIDI are then rendered to audio with FluidSynth\footnote{\url{https://github.com/fluidsynth/fluidsynth}} using the FluidR3\_GM general MIDI soundfont, so that the reference and the edited signal share an identical instrument mapping and any measured difference is attributable to the edit rather than to synthesis.

\begin{table*}[ht!]
  \centering
  \scalebox{0.82}{
    \begin{tabular}{lc p{10.5cm}}
    \toprule
    \textbf{Editing Operation} & \textbf{Target Facet} & \textbf{Description} \\
    \midrule
    +7 semitones (Perfect Fifth) & Harmony & Transpose every pitch up by a perfect fifth. The tonic moves one step clockwise on the circle of fifths, while note timing, sectional layout, and melodic shape are unchanged. \\
    -3 semitones (Minor Third) & Harmony & Transpose every pitch down by a minor third, moving the tonic three steps along the circle of fifths while preserving all non-harmonic facets. \\
    ABC$\rightarrow$AAA & Structure & Replace the second and third sections with repetitions of the first, collapsing a three-part form into a single repeated block. \\
    ABC$\rightarrow$ABA & Structure & Reorder the sections into a ternary reprise form, changing the sectional layout while reusing the original material. \\
    Tempo shifted +50\% & Rhythm \& Meter & Scale the global tempo by a factor of 1.5, compressing all inter-onset intervals while leaving pitch content, harmony, and sectional form intact. \\
    All beats displaced by a constant 150 ms & Rhythm \& Meter & Apply the same timing offset to every beat, introducing a fixed phase shift without changing tempo, note content, or note durations. \\
    Interval direction flipped for the main melody & Melody \& Motif & Invert the sign of every melodic interval in the main melody, reversing the contour direction while keeping interval magnitudes and rhythm. \\
    Climax phrase of the main melody inverted & Melody \& Motif & Melodically invert only the climax phrase, a localized contour change confined to a single salient region. \\
    Grand Piano $\rightarrow$ Electric Piano & Timbre & Set the general MIDI program number of every pitched track to Grand Piano and then change it into Electric
Piano, leaving the percussion track and all note content, timing, and sectional
layout unchanged. \\
    Grand Piano $\rightarrow$ Acoustic Guitar & Timbre & Set the general MIDI program number of every pitched track to Grand Piano and then change it into Acoustic Guitar, leaving the percussion track and all note content, timing, and sectional
layout unchanged. \\
    \bottomrule
    \end{tabular}}
    \caption{Definition of the ten editing operations used for objective validation. Each operation targets one single music facet and leaves the remaining facets unchanged.}
  \label{tab:edit_definitions}%
\end{table*}%

\subsection{Derivation of the Expected Values}
\label{app:objective_expected}
The \textit{Expected} entries in Table~\ref{tab:objective_validation} are derived from the definition of each metric under a known edit rather than chosen post hoc. We give the derivation per facet.

\textbf{Harmonic Transpositions.}
A perfect fifth advances the tonic by one position on the circle of fifths, so Equation \ref{eq1} gives $\tfrac{1}{6}\min(1,11) \approx 0.167$.
A downward minor third advances it by three positions, giving $\tfrac{3}{6} = 0.5$ whenever the key estimate is correct for both signals; the stated upper end of $0.67$ accommodates the cases in which the Krumhansl--Schmuckler estimate lands on an adjacent key, which costs one further step.
Both transpositions rotate the pitch-class distribution without changing its shape, so ChromaSim and ChromaDTWS should fall below 1 while remaining well above 0.
Neither operation touches onset times or sectional layout, so the rhythmic and structural metrics should stay near their preservation ceiling.

\textbf{Structural Reorderings.}
Reordering or duplicating sections reuses the original material, so the global key, the pitch-class distribution, and the global tempo are unchanged, and the harmonic metrics together with $\Delta$\text{BPM} should remain at their preservation ceiling.
Beat positions, however, are displaced past the first reordered boundary, so BeatF and IG should fall below 1 even though no local rhythmic content is altered.

\textbf{Rhythmic Perturbations.}
A +50\% tempo scaling multiplies the underlying pulse rate by 1.5, which is not a metric-octave relation and therefore survives the folding in Equation \ref{eq4}, so $\Delta$BPM should be large, and both BeatF and IG should fall.
A constant offset leaves the pulse rate untouched, so $\Delta$BPM will be near 0.
Because 150ms exceeds the $\pm$ 70ms matching tolerance,
BeatF will collapse toward 0, while IG will remain moderate, since a fixed offset concentrates the timing error at a single phase rather than spreading it uniformly.

\textbf{Melodic Transformations.}
Contour inversion is applied to the main melody track only, which accounts for a median of 9\% of the notes in the mix. It is a chromatic inversion and therefore does alter the melody's pitch-class content, but the harmonic, rhythmic, and structural metrics read the full polyphonic mix and are dominated by the unchanged accompaniment, so they should remain high.
ContourDTWS will fall, since inversion changes the pitch-class trajectory of the melody.
MotifRec will fall, since inverted interval 3-grams are distinct tokens from the originals.
The localized climax inversion perturbs only a small fraction of the interval sequence, so it will move both melodic metrics less than the global interval flip.

\section{Human Listening Questionnaire}
\label{app:human}
We design a questionnaire to collect human auditory perception of the tailored editing settings\footnote{\url{https://github.com/Yashvishe13/MuseCPEval/blob/main/questionnaire.pdf}}.
The questionnaire comprises paired edited music clips, organized as 4 questions per music facet.
Each question presents an original music clip together with two edited versions of that clip.
To validate the metric’s ability to capture musical nuances that align with human auditory perception, the two edited clips are produced by applying the same facet-targeted operation at two different strengths.
We ask the participants to compare the two edited clips and decide which one is farther from the original one with respect to one specific facet.
We also construct samples for a \textit{sanity check} to filter out low-quality responses.
Participants are anonymous and have diverse music backgrounds.
We received 33 questionnaire responses and retained the 11 that passed our sanity checks for the final analysis in Section \ref{sec:human_listening}.



\end{document}